\documentclass[fleqn,10pt]{wlscirep}
\title{Completely scrambled memory for quantum superposition}

\author[1,*]{Tetsuya Mukai}
\affil[1]{NTT Basic Research Laboratories, NTT Corporation, 3-1, Morinosato-Wakamiya, Atsugi, Kanagawa 243-0198, Japan}

\affil[*]{mukai.tetsuya@lab.ntt.co.jp}

\begin{abstract}
Although constructing a quantum computation device with multiple qubits is arguably a difficult task, several seconds of coherence time with tens of thousands of quantum particles has been demonstrated with a trapped atomic ensemble. As a practical application, a security-enhanced quantum state memory using atoms has been demonstrated. It was shown that the quantum superposition preserved in an atomic ensemble was scrambled and faithfully descrambled; however, the scrambled phase ambiguity remained at 50~\%. To overcome this problem, we propose and demonstrate a scheme that achieves 100~\% phase ambiguity without introducing an extra Ramsey interferometer. Moreover, this scheme can be used as a direct application to keep the choice between two values secret without falsification.
\end{abstract}
\begin{document}

\flushbottom
\maketitle
\thispagestyle{empty}

\section*{Introduction}

A two-level quantum system is a useful resource for quantum information science. We can find two-level quantum systems in a wide range of hardware, and a neutral atom is one example. The advantage of an atomic system is its long coherence time, which is realized because of its complete isolation from the environment. Moreover, we can obtain a large number of atoms that have exactly the same properties, and thus can expect good scalability.

A widely used and important application of an atomic system is a quantum computation device that outperforms a conventional classical computer. In one feasible quantum gate operation approach, a physical qubit is constituted with a trapped single atom that is capable of achieving independent control of the state. Moreover, multiple atomic qubits must be cooperatively controlled to generate entangled states. This type of approach is technically challenging, but work is progressing \cite{Isenhower_2010, Wilk_2010, Wang_2015, Lester_2015, Kaufman_2015, Bernien_2017}.
  
On the other hand, it has been demonstrated that a few tens of thousands of atoms trapped in a potential can be coherently controlled for more than a second by an externally applied electromagnetic field \cite{Deutsch_2010, Bernon_2013}. In this case, each trapped atom is in practically the same quantum state and experiences a uniform time evolution; therefore, the atomic ensemble represents a single quantum state. Although this system cannot handle entangled states with more than two qubits, a trapped atomic ensemble is useful as a memory for a quantum state. A significant merit of using this memory relates to the identification of the superposition state. In quantum state identification, we usually need to average multiple results of single-qubit measurements obtained after the same operation. In contrast, with a trapped atomic ensemble, as long as the influence of the atom-atom interaction is negligible, a single distribution measurement in the ground or excited state is sufficient to identify the quantum state. With this state identification technique, we can facilitate the efficiency verification of a gate-operation protocol over a qubit.
  
As an example of protocol verification, in our previous report we demonstrated a security-enhanced memory for quantum superposition \cite{Mukai_2017}. In that demonstration, ambiguous phase fluctuation was introduced by a Ramsey interferometer that had a random phase difference relative to another Ramsey interferometer that recorded the quantum state. However, in this scheme, as long as two independent Ramsey interferometers are used and a half ground and half excited state is recorded, the maximum phase ambiguity that can be introduced is limited to as little as 50~\%, and the frequency of the Ramsey flop is inferred from the measurement results.

In this article, we study the phase ambiguity of a security-enhanced quantum memory designed to overcome the above problem and demonstrate 100~\% phase ambiguity with an atomic ensemble trapped below a persistent supercurrent atom chip \cite{Mukai_2007, ChipBEC_2014}. Moreover, we propose a practical application for recording a choice between two values, e.g., yes or no, and keeping it secret without falsification.

\section*{Results}
\subsection*{Phase ambiguity of security-enhanced quantum memory}
An atomic ensemble can be used as a memory for quantum superposition by using Ramsey interferometry (see Methods). The security of the memory is enhanced by employing another Ramsey interferometer as described in detail in our previous report \cite{Mukai_2017}. For consistency, we employ the same abbreviations - WRI (write-read Ramsey interferometer) and SRI (scramble-retrieve Ramsey interferometer). In addition, we call the first and second pulses of the WRI (SRI) {\it write} and {\it read} ({\it scramble} and {\it retrieve}) pulses, respectively.

\begin{figure*}[thbp]
\begin{center}
\includegraphics[width=16cm]{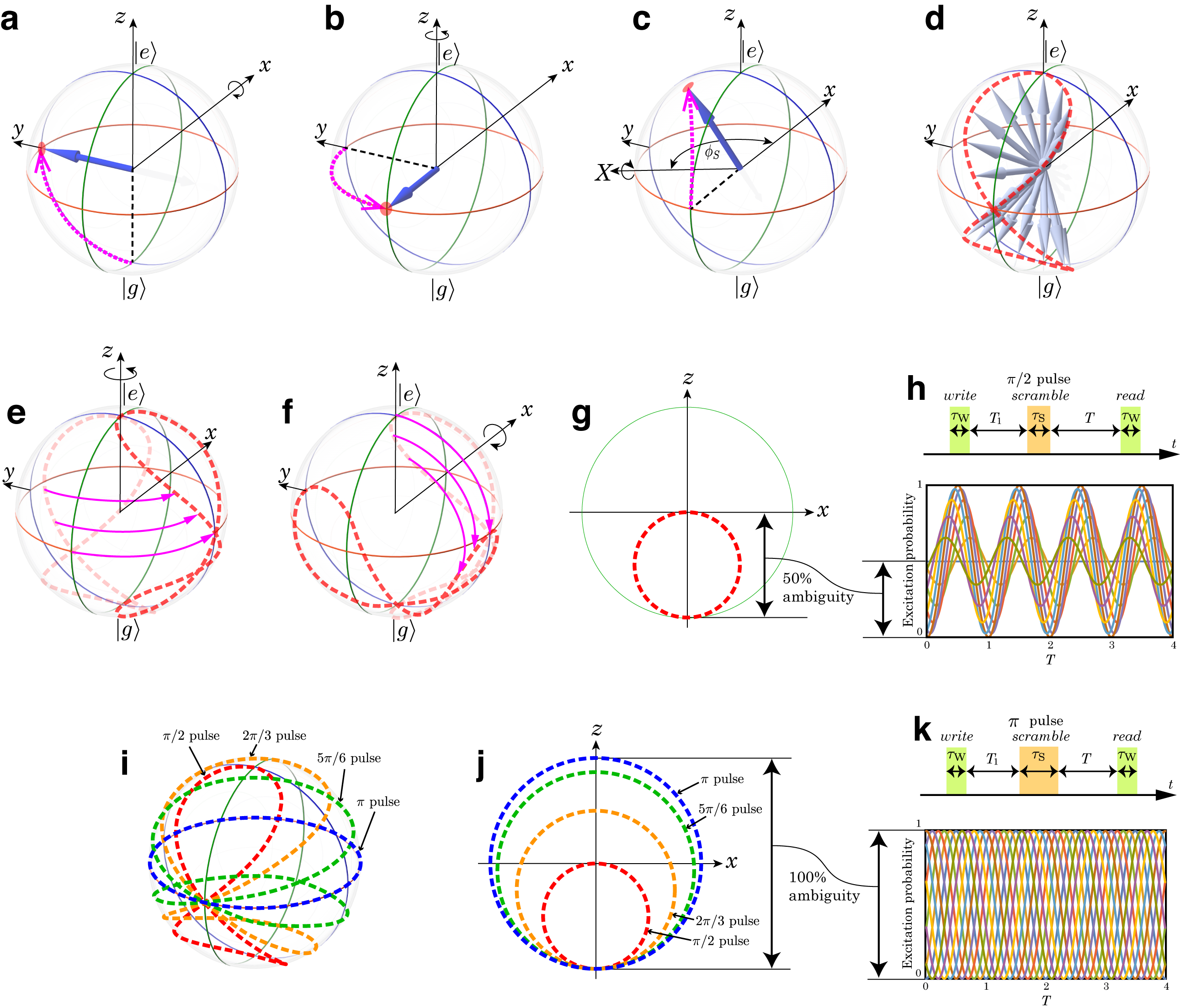}
\caption{Bloch vector analysis. ({\bf a}) $\pi/2$ rotation of the Bloch vector around the $x$-axis. ({\bf b}) Time evolution of the Bloch vector during the time interval. ({\bf c}) Rotation of the Bloch vector around the $X$-axis caused by irradiating the {\it scramble} pulse. ({\bf d}) The dashed curve represents a stochastic distribution of the Bloch vector (SDBV) after irradiating a {\it scramble} pulse. ({\bf e}) Time evolution of the SDBV. ({\bf f}) Rotation of the SDBV caused by irradiating the {\it read} pulse. ({\bf g}) SDBV projected on the $xz$-plane. ({\bf h}) Pulse sequence and excitation probability of Ramsey flop scrambled with a $\pi/2$ pulse. ({\bf i}) Pulse area dependence of an SDBV on a Bloch sphere. ({\bf j}) Pulse area dependence of an SDBV projected on the $xz$-plane. ({\bf k}) Pulse sequence and excitation probability of a Ramsey flop scrambled with a $\pi$ pulse. In ({\bf h}) and ({\bf k}) several excitation probabilities are calculated with a phase difference $\phi_{S}\in[0, 2\pi]$ and plotted in different colors. The horizontal axis $T$ in ({\bf h}) and ({\bf k}) is normalized with $2\pi/\delta_{\rm W}$, where $\delta_{\rm W}$ represents Raman detuning. With these plots, ({\bf h}) and ({\bf k}), we can estimate the range of shot to shot fluctuations of the excitation probability expected to be observed in the Ramsey flop measurement.}
\label{Bloch_vector_analysis}
\end{center}
\end{figure*}

In the demonstration described in the previous report, we employed two Ramsey interferometers with two $\pi/2$ pulses, and the scrambled phase ambiguity remained at 50~\%. To understand the mechanism of the limited phase ambiguity, we track the Bloch vector of an atomic ensemble through the memory scrambling process. First, a superposition state is recorded with a {\it write} pulse irradiating on an atomic ensemble that has been initialized in the ground state, and the Bloch vector rotates $\pi/2$ radians around the $x$-axis (Fig.\ref{Bloch_vector_analysis}-a). After irradiating the {\it write} pulse, the atomic ensemble experiences a time evolution during a time interval $T_{1}$, and the Bloch vector rotates around the $z$-axis (Fig.\ref{Bloch_vector_analysis}-b). Next, a {\it scramble} pulse is irradiated on the atomic cloud to encrypt the recorded superposition state. By irradiating the {\it scramble} pulse, the Bloch vector rotates around the $X$-axis, forming an angle $\phi_{S}$ with the $x$-axis on the $xy$-plane. Figure \ref{Bloch_vector_analysis}-c shows the rotation of the Bloch vector in the scrambling process when the phase evolution during $T_{1}$ equals $\pi/2$ radians and $\phi_{S}$ equals $2\pi/3$ radians. Depending on the random phase difference between the WRI and SRI, $\phi_{S}$ stochastically takes a value between $0$ and $2\pi$ radians. Therefore, the Bloch vector is stochastically transferred to a point on a curve on the Bloch sphere. The dashed curve in Fig.\ref{Bloch_vector_analysis}-d shows the calculated stochastic distribution of the Bloch vector (SDBV) after irradiating a {\it scramble} pulse. 

We prepare for the state distribution measurement by irradiating a {\it read} pulse after the SDBV has rotated around the $z$-axis in a time interval $T$ (Fig.\ref{Bloch_vector_analysis}-e). By irradiating the {\it read} pulse, the SDBV rotates $\pi/2$ radians around the $x$-axis (Fig.\ref{Bloch_vector_analysis}-f). A projection of the SDBV onto the $xz$-plane is plotted in Fig.\ref{Bloch_vector_analysis}-g, which represents a circle 0.5 in diameter. 
This is an important outcome in terms of understanding the ambiguity of the excitation probability of an atomic cloud. The SDBV along the z-axis is the reason for the range of shot to shot fluctuations in the excitation probability when measuring the scrambled state. Depending on the time interval $T$, the projected SDBV circle rotates around the origin of the $xz$-plane, but the size of the distribution along the $z$-axis is equal to the diameter of the SDBV circle regardless of the $T$ value. Therefore, we can understand the 50~\% ambiguity with respect to the excitation probability of the Ramsey flop regardless of the $T$ value (Fig.\ref{Bloch_vector_analysis}-h).

To increase the ambiguity of the excitation probability of the scrambled state, the SDBV range along the $z$-axis should be extended. This can be accomplished by changing the area of the {\it scramble} pulse. Figure \ref{Bloch_vector_analysis}-i shows SDBVs calculated with different {\it scramble} pulse areas. The SDBVs projected onto the $xz$-plane after a $\pi/2$ radians phase evolution and irradiating the {\it read} pulse are plotted in Fig.\ref{Bloch_vector_analysis}-j. The SDBV shows a clear dependence on the scramble pulse area, and it becomes the equator of the Bloch sphere when the {\it scramble} pulse area equals $\pi$. Therefore, we can expect 100~\% ambiguity of the excitation probability of the Ramsey flop with $\pi$ pulse scrambling as plotted in Fig.\ref{Bloch_vector_analysis}-k.

\subsection*{Experimental demonstration}
\begin{figure}[thbp]
\begin{center}
\includegraphics[width=16cm]{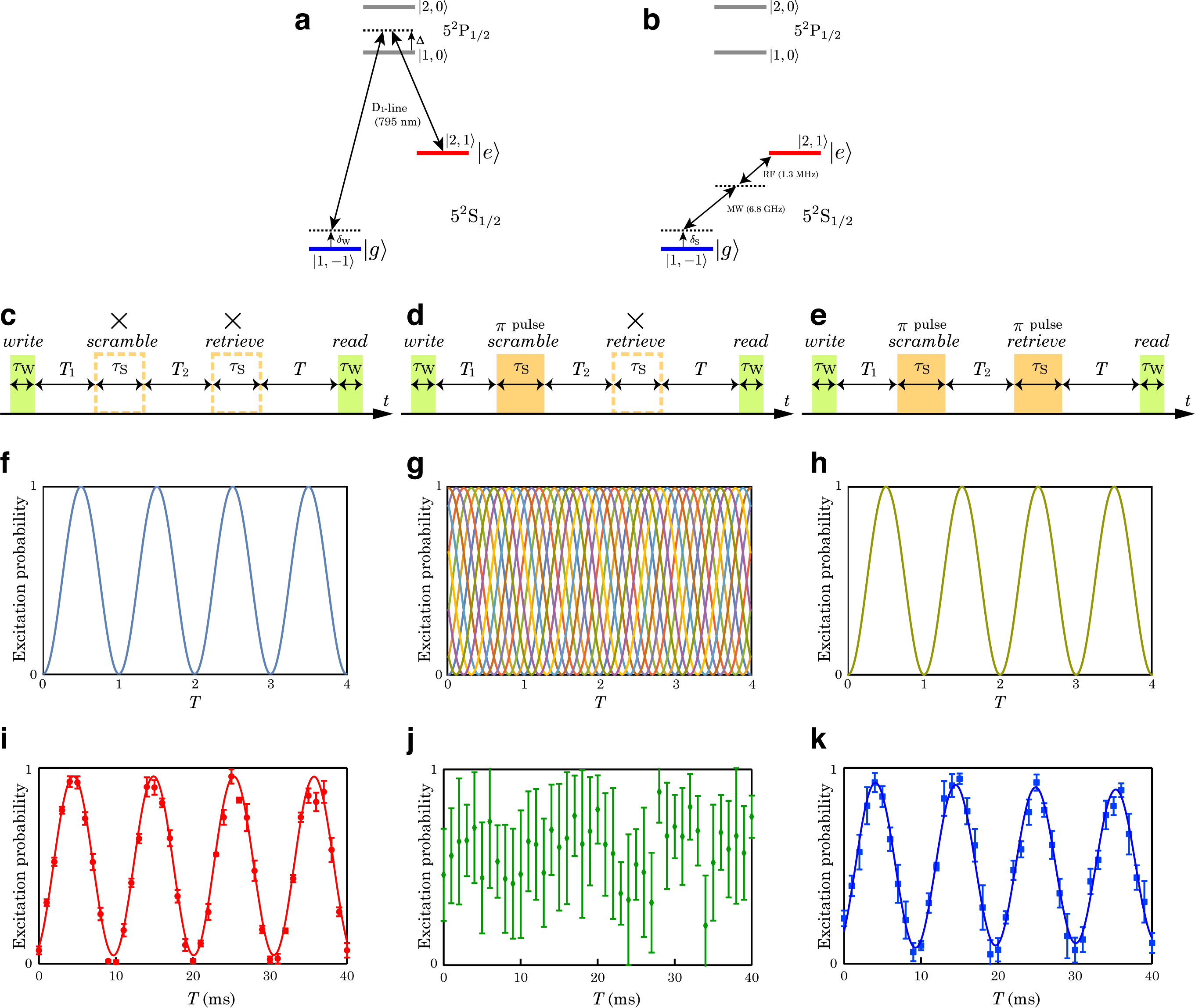}
\caption{Complete scramble and retrieve operation and results. ({\bf a}) D$_{1}$-line stimulated Raman transition for the WRI. ({\bf b}) MW-RF transition for the SRI. ({\bf c}) Normal Ramsey pulse sequence. ({\bf d}) Scrambled Ramsey pulse sequence. ({\bf e}) Scrambled and retrieved Ramsey pulse sequence. ({\bf f}) Calculated normal Ramsey flop. ({\bf g}) Calculated scrambled Ramsey flop. ({\bf h}) Calculated scrambled and retrieved Ramsey flop. ({\bf i}) Experimentally obtained normal Ramsey flop. ({\bf j}) Experimentally obtained scrambled Ramsey flop. ({\bf k}) Experimentally obtained scrambled and retrieved Ramsey flop. The horizontal axis $T$ in ({\bf f}), ({\bf g)}, and ({\bf h}) is normalized with $2\pi/\delta_{\rm W}$, where $\delta_{\rm W}$ represents Raman detuning of the WRI. Several curves with a phase difference $\phi_{S}\in[0, 2\pi]$ are plotted in both ({\bf g}) and ({\bf h}) plots in different colors. The lines in both ({\bf i}) and ({\bf k}) are damped sinusoidal fit to the data.}
\label{Experimental_data}
\end{center}
\end{figure}

The demonstration was performed with $^{87}{\rm Rb}$ Bose-Einstein condensate trapped below a persistent supercurrent atom chip \cite{Mukai_2007, ChipBEC_2014}. The experimental setup was the same as that in our previous work \cite{Mukai_2017}, but we improved the coherence time by suppressing radio frequency noise. We used the spin states $| F, m_{\rm F} \rangle = |1, -1\rangle$ and $| 2, 1\rangle$ of $5^{2}{\rm S}_{1/2}$ for the ground and excited states of the Ramsey interferometer, and we denote them as $|g\rangle$ and $|e\rangle$, respectively. These two states are coupled through a two-photon stimulated Raman transition realized with a pair of ${\rm D}_{1}$-line lasers (Fig.\ref{Experimental_data}-a) or a pair consisting of microwave (MW) and radio frequency (RF) fields (Fig.\ref{Experimental_data}-b). The pair of ${\rm D}_{1}$-line lasers was employed as a coupling field for the WRI, and the pair consisting of MW and RF fields was similarly employed for the SRI. The roles of the WRI and SRI can be swapped without making any practical difference. The coherence time of both Ramsey interferometers was of the order of seconds, while the pulse duration time for a $\pi/2$ pulse was of sub-millisecond order. The {\it retrieve} pulse was irradiated when the SRI experienced a $(2m+1) \pi$ radians phase evolution, where $m$ is an integer. The parameters used in the experiment are summarized in Table \ref{tab:table1}.

\begin{table}[ht]
\centering
\begin{tabular}{|l|l|l|}
\hline
parameter & value\\
\hline
WRI pulse duration time & $\tau_{W}$ = $0.45~{\rm ms}$\\
WRI Raman detuning & $\delta_{W}$ = $2\pi \times 100~{\rm Hz}$\\
SRI pulse duration time & $\tau_{S}$ = $2.75~{\rm ms}$\\
SRI Raman detuning & $\delta_{S}$ = $2\pi \times 100~{\rm Hz}$\\
{\it write}-{\it scramble} time interval & $T_{1}$ = $5~{\rm ms}$\\
{\it scramble}-{\it retrieve} time interval & $T_{2}$ = $5~{\rm ms}$\\
\hline
\end{tabular}
\caption{\label{tab:table1}Parameters used in the experiment}
\end{table}

To demonstrate complete scrambling, we employed $\pi$ pulse scrambling and measured three Ramsey flops with different pulse application sequences, i.e., normal, scrambled, and retrieved Ramsey flops. The pulse sequences are shown in Fig.\ref{Experimental_data}-c, -d, and -e. The Ramsey flops calculated with each condition are plotted in Fig.\ref{Experimental_data}-f, -g, and -h, respectively. In these calculations, we neglected the pulse duration time compared to the time interval $T$. For both the scrambled (Fig.\ref{Experimental_data}-g) and retrieved (Fig.\ref{Experimental_data}-h) Ramsey flops, several curves with a phase difference between the WRI and SRI ($\phi_{S}\in [0, 2\pi]$) are plotted in different colors, however, there is no difference between the curves for the retrieved Ramsey flop in Fig.\ref{Experimental_data}-h. With these calculations we can estimate the range of the shot to shot fluctuations of the excitation probability that we can expect to observe in the Ramsey flop measurement.

The Ramsey flops obtained experimentally for each condition are plotted in Fig.\ref{Experimental_data}-i, -j, and -k, respectively. We performed five measurements and averaged the results to obtain a single data point with a standard deviation. Before scrambling, a clear Ramsey flop was achieved as plotted in Fig.\ref{Experimental_data}-i. When the state was scrambled, the measured standard deviation of the excitation probability was sufficiently large regardless of the $T$ value, and no information remained about the Ramsey flop as shown in Fig.\ref{Experimental_data}-j; i.e., the Ramsey flop was completely scrambled. The scrambled Ramsey flop was faithfully descrambled by irradiating the {\it retrieve} pulse as shown in Fig.\ref{Experimental_data}-k. The slight increase in the standard deviation was probably caused by the inaccuracy of the {\it scramble} and {\it retrieve} pulses.

\section*{Discussion}
The crux of the phase ambiguity achieved with this method is the SDBV after irradiating the {\it scramble} pulse. 100~\% phase ambiguity is achieved when the SDBV becomes the equator of the Bloch sphere. This can be achieved with $\pi$ pulse scrambling when an equal superposition of the ground and excited states is recorded. However, $\pi$-pulse scrambling does not always result in the maximum phase ambiguity when the other superposition state is recorded; e.g. when an excited state $|e\rangle$ is recorded, $\pi/2$ pulse rather than $\pi$ pulse  scrambling results in 100~\% phase ambiguity. When a general superposition is recorded, it is basically impossible to make the SDBV the equator of the Bloch sphere. We still have room to optimize the area of the {\it scramble} pulse depending on the recorded superposition state.

\begin{figure}[thbp]
\begin{center}
\includegraphics[width=16cm]{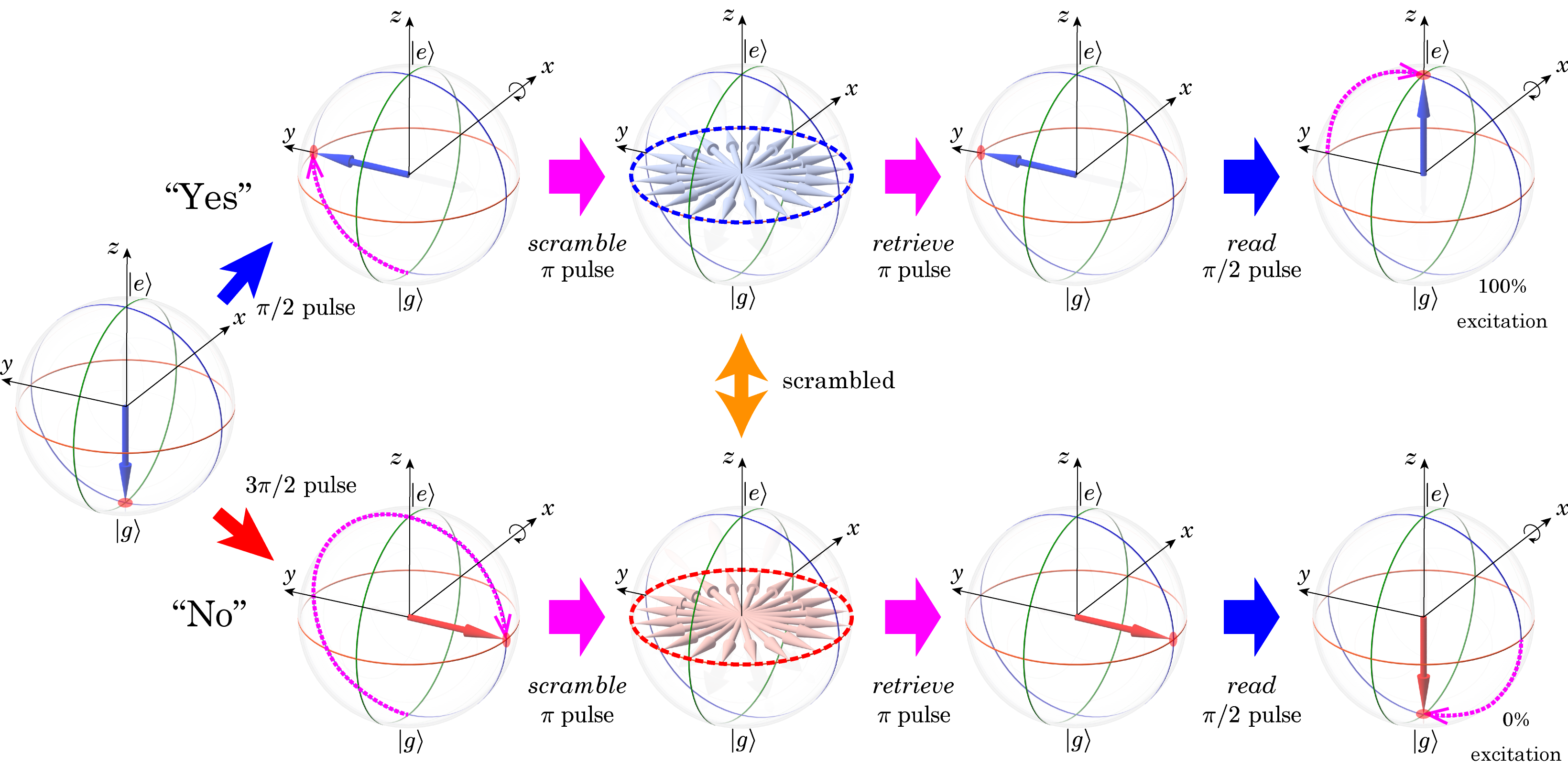}
\caption{Bloch vector representation of the scheme for securely recording a choice between two values. While the {\it write} pulse area ($\pi/2$ or $3\pi/2$) depends on the choice ``yes" or ``no", the other pulse area is independent of the choice. Once scrambled, the Bloch vector stochastically distributes a point on the equator of the Bloch sphere in both cases, and we cannot infer the choice without descrambling using a suitable {\it retrieve} pulse.}
\label{secure_recording_scheme}
\end{center}
\end{figure}

Although there remains a requirement for the state to be recorded, the demonstrated 100~\% phase ambiguity promises useful applications in security enhancement. As an example, we propose a scheme for recording a choice between two values, e.g., yes or no, and keeping it secret without falsification (Fig.\ref{secure_recording_scheme}). 

We assume that the choice ``yes" or ``no" is mapped to a superposition state $|+\rangle = \frac{1}{\sqrt{2}}(|g\rangle+|e\rangle)$ or $|-\rangle = \frac{1}{\sqrt{2}}(|g\rangle-|e\rangle)$ by employing a {\it write} pulse with a $\pi/2$ or $3\pi/2$ pulse, respectively. In the $\pi$ pulse scrambling process, the Bloch vector of $|+\rangle$ is stochastically transferred to a point on the equator of the Bloch sphere and that of $|-\rangle$ is stochastically transferred to a point on the same equator. At this point, the initial choice of ``yes" or ``no" cannot be identified from the measurement of the Bloch vector, and falsification is practically impossible.

The preparation for the state readout is performed as follows. The memory is descrambled by irradiating a {\it retrieve} pulse when the SRI phase evolution is equal to $(2l+1) \pi$ radians, where $l$ is an integer. Then a {\it read} pulse, which has a $\pi/2$ pulse area irrespective of the choice at the writing process, is irradiated when the WRI phase evolution is equal to $2k\pi$ radians, where $k$ is an integer. Then we can understand that the choice written on the memory was ``yes" or ``no" by obtaining 100~\% or 0~\% excitation, respectively.

In summary, we investigated the phase ambiguity of a security-enhanced quantum memory and made it clear that the ambiguity is dependent on the area of the {\it scramble} pulse. Moreover, we proposed and experimentally demonstrated a scheme designed to achieve 100~\% phase ambiguity, which is useful for recording a choice between two values without falsification. This technique can be applied to a wide range of two-level quantum systems that can be controlled with Ramsey interferometric methods.

\section*{Methods} \label{Methods}
\subsection*{Atomic ensemble memory for quantum superposition}
An atomic ensemble is initially prepared in the ground state, and a quantum state is recorded by irradiating a {\it write} pulse. The recorded state is verified by irradiating a {\it read} pulse after a specific time interval. As long as the {\it write} and {\it read} pulses have the same pulse area, and the phase evolution during the time interval is equal to $(2n + 1)\pi$ radians, where n is an integer, faithful recording can be confirmed by measuring the distribution in the initial state.


\section*{Acknowledgements}

This work received support from the Japan Science and Technology Agency (JST) through its Core Research for Evolutional Science and Technology (CREST), JPMJCR1671.

\section*{Author Contributions}

T.M. conceived and conducted the experiment, analyzed the results, and wrote the manuscript.

\end{document}